\documentclass[aps,chaos,preprint,groupedaddress,showpacs,showkeys]{revtex4}

\usepackage{graphicx}
\usepackage{dcolumn}
\usepackage{bm}
\usepackage{amssymb}

\graphicspath{{Fig/}}

\begin{document}

\title{On-Off Intermittency in Time Series of Spontaneous Paroxysmal Activity \protect\\in Rats with Genetic Absence Epilepsy}

\author{A.E.~Hramov}
\email{aeh@cas.ssu.runnet.ru}
\author{A.A.~Koronovskii}
\email{alkor@cas.ssu.runnet.ru}
%
\affiliation{Faculty of Nonlinear Processes, Saratov State
University, Astrakhanskaya str., 83, Saratov, 410012, Russia}
%
\author{I.S.~Midzyanovskaya}
\email{miinn@yandex.ru}
\author{E.~Sitnikova}
\email{jenia-s@mail.ru}
\affiliation{Institute of the Higher Nervous Activity and
Neurophysiology of Russian Academy of Sciences,
Butlerova str., 5A, Moscow, 117485, Russia}
\author{C.M.~van Rijn}
\email{rijn@nici.ru.nl}
\affiliation{NICI--Biological Psychology, Radboud University Nijmegen, PO 9104,
6500 HE Nijmegen, The Netherlands}
\date{\today}

\begin{abstract}
\noindent In the present paper we report on the on-off intermittency
phenomena observed in time series of spontaneous paroxysmal
activity in rats with genetic absence epilepsy. The method to
register and analyze the electroencephalogram with the help of
continuous wavelet transform is also suggested.
\end{abstract}


\maketitle

{\bf Dynamic behavior of complex neuronal ensembles is a topic
comprising a streamline of current researches worldwide. In this
article we study the behavior manifested by epileptic brain, in
the case of spontaneous non-convulsive paroxysmal activity. For
this purpose we analyzed archived long-term recording of
paroxysmal activity in animals genetically susceptible to absence
epilepsy, namely WAG/Rij rats. We first report that the brain
activity alternated between normal states and epilepsy paroxysms
is the on-off intermittency phenomenon which has been observed and
studied earlier in the different nonlinear systems.}

\section{Introduction}

Dynamic behavior of complex neuronal ensembles is a topic
comprising a streamline of current researches worldwide. Brain
considered as a composition of neuronal ensemble is a challenging
subject for nonlinear dynamics. Revelation of intrinsic dynamic
regularities of brain, under normal and pathological conditions,
can shed more light upon origination and evolution of brain
diseases~\cite{Tass:2003_NeuroSynchro}.

In computational neuroscience one can use theoretical approaches,
based on models of single neurons~\cite{Velarde:2002_NeuronModel},
which can be further linked with each
other~\cite{Nekorkin:2000_Neuro}. Lattices and chains of such
elements, modeling particular brain regions and
nuclei~\cite{Kazantsev:2003_NeuronLattice}, are the next advance
in theoretical approximation of cunning brain anatomy. Moreover,
the networks of nonlinear elements~\cite{Chavez:2005_Networks_PRL,
Hwang:2005_Newtwork_PRL} are also the interesting objects to be
studied. Unambiguously, results obtained in computational
neuroscience should be compared with facts yielded by experimental
neuroscience.

Application of methods developed in modern nonlinear dynamics to
experimental time series is an additional approach to
understanding of brain functions \cite{Tass:1998_NeuroSynchro,
Rosenblum:2001_HandbookBiologicalPhysics,
Boccaletti:2005_Chaos_TSS}. Particularly, estimation of functional
interaction between brain regions by the means of quantification
of a directionality and strength of functional coupling calculated
from local field potentials (i.e., electroencephalogram, EEG)
\cite{Smirnov:2005_CouplingDirection} may be noted.

The present work was aimed to diagnose the type of dynamic
behavior manifested by epileptic brain, in the case of spontaneous
non-convulsive paroxysmal activity. For this purpose we analyzed
archived long-term records of paroxysmal activity in animals
genetically susceptible to absence epilepsy, namely WAG/Rij
rats~\cite{Luijtelaar:1986}. We report that this paroxysmal
activity is on-off intermittency~\cite{Platt:1993_intermittency,
Heagy:1994_intermittency,Boccaletti:2000_IntermitLagSynchro}.

The structure of the paper is the following. In
section~\ref{sct:Methods} we describe the experimental setup and
the methods to register and analyze the electroencephalogram in
animals. The obtained results are given in
section~\ref{sct:Results}. The final conclusion is presented in
section~\ref{sct:Conclusion}.

\section{Methods}
\label{sct:Methods}

\begin{figure}[p]
\centerline{\scalebox{0.6}{\includegraphics{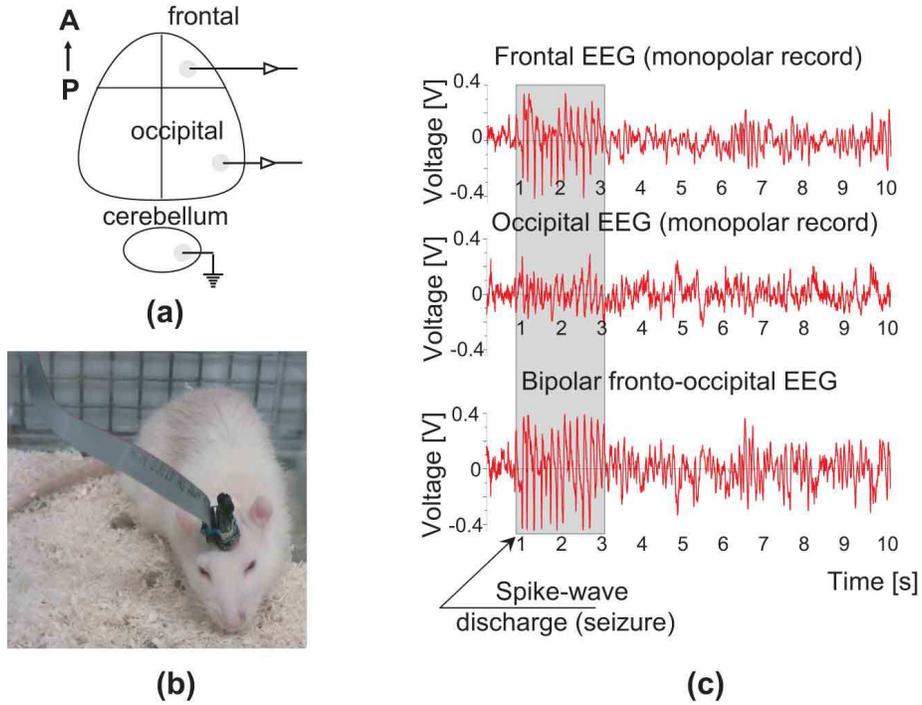}}}
\caption{(\textit{a}) Rat with the implanted electrodes in the
experimental camera during the electroencephalogram registration.
(\textit{b}) The typical fragment of the electroencephalogram of
the rat registered from the frontal area of the brain cortex. The
time interval corresponding to the epilepsy paroxysm is shown by
the gray rectangle} \label{fgr:Rats}
\end{figure}

The animals (6 females, weighing 230--250 g, 5--6 months old and 5
males, weighing 250--300g, 5--6 months old) were chronically
implanted with electrodes (stainless screws), let to post-surgery
recovering and habituation to experimental camera
(see~\ref{fgr:Rats},\,\textit{a}). Paroxysmal activity was
quantified from EEG registered in freely moving animals by means
of automatic routine followed by an expert assessment as
described elsewhere~\cite{Luijtelaar:1986}. The duration of EEG
registration varied from 6 hours up to 4 days. An example of EEG
fragment containing background activity of low amplitude
interspersed with absence paroxysms of relatively high voltage is
given in Fig.~\ref{fgr:Rats},\,\textit{b}. Absence paroxysms are
manifested as generalized over the brain cortex synchronous
spike-wave discharges (SWDs) with abrupt onset and
ending~\cite{Luijtelaar:1986}.

Together with the traditional approach mentioned above we have
also used the continuous wavelet
transform~\cite{alkor:2003_WVTBookEng, Daubechies:1992_WVTBook,
Kaiser:1994_Wvt, Torresani:1995_WVT} to select the fragments
corresponding to the absence epilepsy paroxysms from recorded
electroencephalograms because the continuous wavelet transform is
the powerful tool for the analysis of nonlinear dynamical system
behavior. In particular, the continuous wavelet analysis has been
used for the detection of synchronization of chaotic oscillations
in the brain~\cite{Lachaux:2000_WVTSynchro,
Lachaux:2002_BrainCoherence, Quyen:2001_WVTvsHilbert},
cardiovascular human system~\cite{Hramov:2006_Prosachivanie},
chaotic laser array~\cite{DeShazer:2001_WVT_LaserArray}, etc.
Continuous wavelet transform has also been used to study the time
scale synchronization~\cite{Hramov:2004_Chaos,
Aeh:2005_SpectralComponents, Aeh:2005_TSS:PhysicaD}. Since the
fragments of electroencephalogram corresponding to the normal
brain dynamics and absence epilepsy paroxysms seem to be quite
different the continuous wavelet transform is the convenient
mathematical tool for the different types of the brain activity to
be distinguished.

Obviously, the brain activity alternated between normal states and
epilepsy paroxysms may be considered as the intermittent behavior.
Therefore, let us term the low-amplitude fragments of
electroencephalogram (corresponding to the ``normal'' brain state)
as the laminar phases and the high-amplitude fragments
(corresponding to the absence epilepsy paroxysms) as the turbulent
ones.

The core idea of the method of the division of time series into
the laminar and turbulent phases by means of the continuous
wavelet transform is the following. The continuous wavelet
transform of chaotic time series $x(t)$ is given by
\begin{equation}
W(s,t_0)=\int\limits_{-\infty}^{+\infty}x(t)\psi^*_{s,t_0}(t)\,dt,
\label{eq:WvtTrans}
\end{equation}
where $\psi_{s,t_0}(t)$ is the wavelet--function related to the
mother--wavelet $\psi_{0}(t)$ as
\begin{equation}
\psi_{s,t_0}(t)=\frac{1}{\sqrt{s}}\psi\left(\frac{t-t_0}{s}\right).
\label{eq:Wvt}
\end{equation}
The time scale $s$ corresponds to the width of the wavelet
function $\psi_{s,t_0}(t)$, and $t_0$ is shift of wavelet along
the time axis, the symbol ``$*$'' in~(\ref{eq:WvtTrans}) denotes
complex conjugation. It should be noted that the time scale $s$ is
usually used instead of the frequency $f$ of Fourier
transformation and can be considered as the quantity inversed to
it.

The Morlet--wavelet~\cite{Grossman:1984_Morlet}
\begin{equation}
\psi_0(\eta)=\frac{1}{\sqrt[4]{\pi}}\exp(j\Omega_0\eta)\exp\left(\frac{-\eta^2}{2}\right)
\label{eq:Morlet}
\end{equation}
has been used as a mother--wavelet function. The choice of
parameter value $\Omega_0=2\pi$ provides the relation ${s\approx
1/f}$ between the time scale $s$ of wavelet transform and
frequency $f$ of Fourier transformation.

\begin{figure}[tb]
\centerline{\scalebox{0.5}{\includegraphics{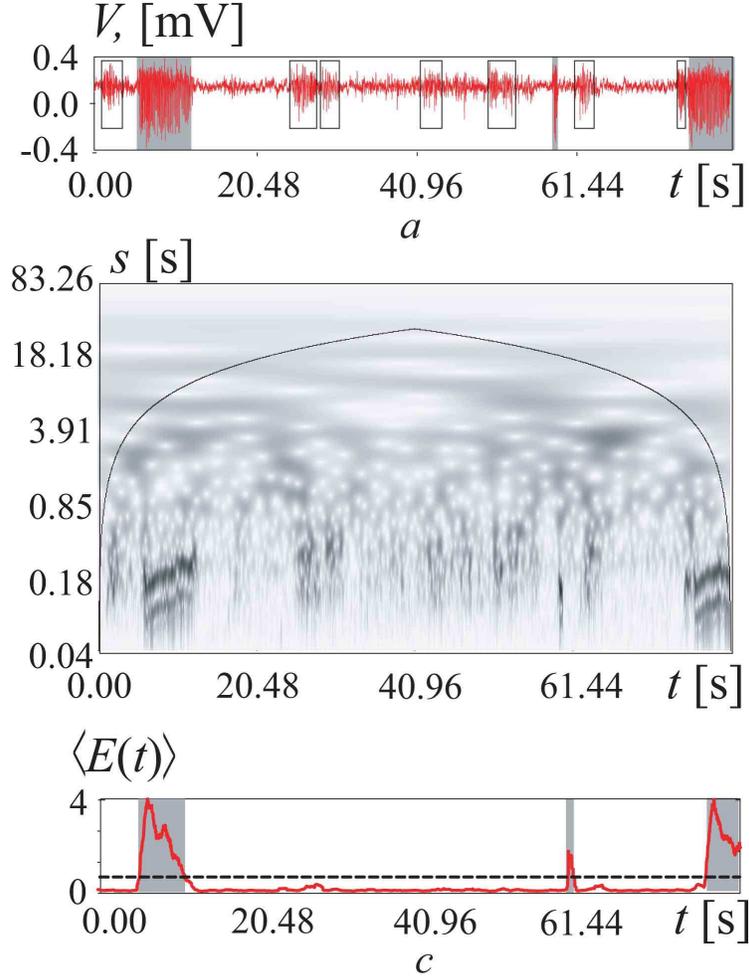}}}
\caption{(\textit{a}) The fragment of electroencephalogram (EEG)
consisting of the laminar phases (normal EEG) alternated with
turbulent phases (epileptic activity, spike-wave discharges,
marked by gray rectangles) corresponding to the epilepsy
paroxysms. (\textit{b}) Projection of the wavelet surface
corresponding to EEG. Time is shown on the abscissa and time scale
is shown on the ordinate. The color intensity is proportional to
the absolute value of the wavelet transform coefficients. The
scales from the right side of the figure show the values of the
coefficients $|W(s, t)|$. The solid line in the figure limits the
area of influence of boundary effects at wavelet spectrum
calculation \cite{alkor:2003_WVTBookEng, Torrence:1998}. Results
of the wavelet spectrum calculation are authentic only below solid
line. (\textit{c}) The dependence of the energy $\langle
E(t)\rangle$ averaged over the characteristic time scale range
$(s_1,s_2)$ on time $t$. The threshold $\Delta E=0.5$ is shown by
the dashed line} \label{fgr:WVTQuanatities}
\end{figure}

\begin{figure}[tb]
\centerline{\scalebox{0.45}{\includegraphics{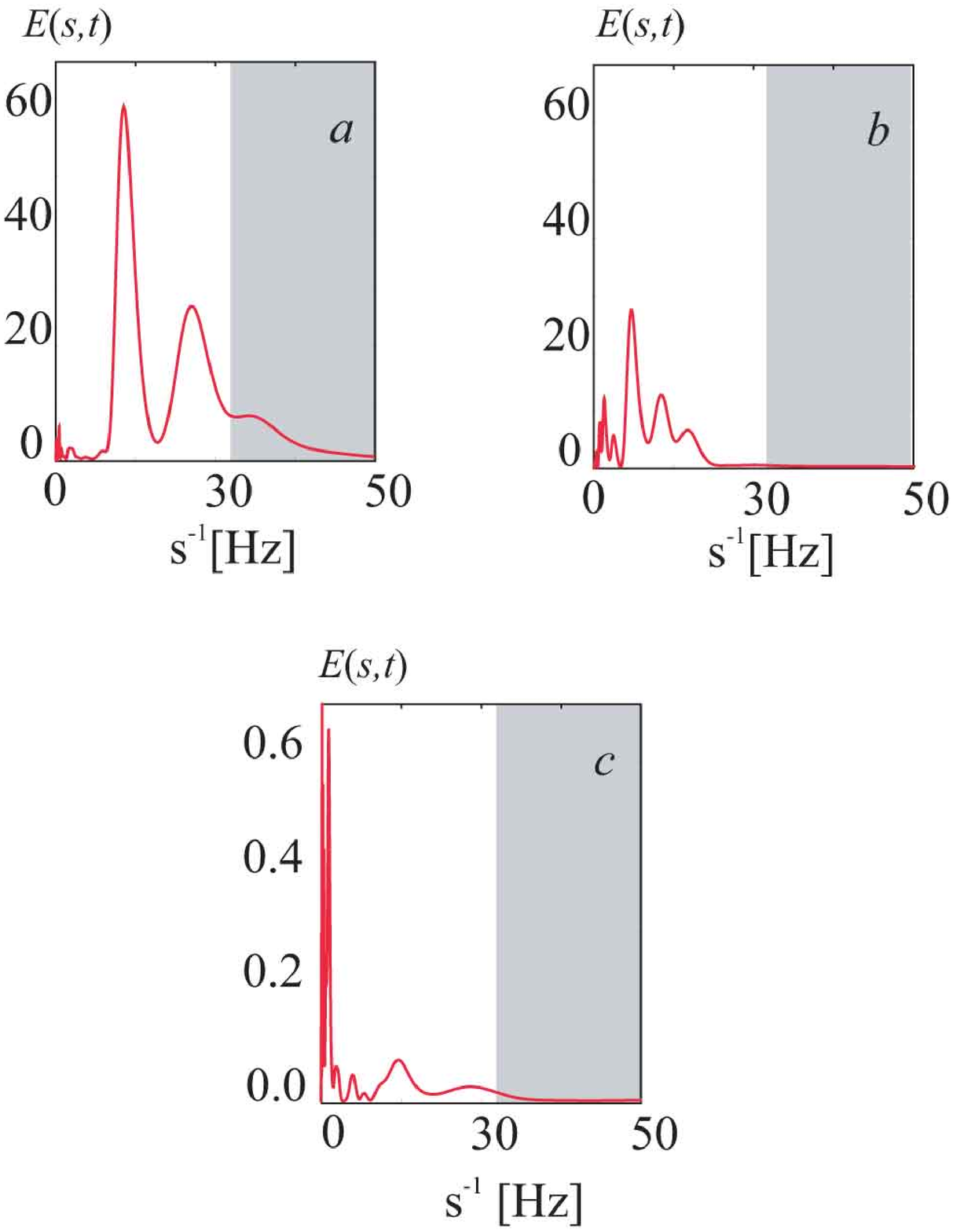}}}
\caption{Typical energy wavelet spectra during turbulent
(\textit{a}) and laminar (\textit{b}) phases in the different
moments of time $t$. Note, scale difference in the axis of
ordinates: laminar phases reveal 100 times less spectral energy in
comparison with turbulent phases} \label{fgr:WVTPowerDistrib}
\end{figure}

The wavelet surface
\begin{equation}
W(s,t_0)=|W(s,t_0)|e^{j\phi_s(t_0)} \label{eq:WVT_Phase}
\end{equation}
describes the system's dynamics on every time scale $s$ at the
moment of time $t_0$. The value of $|W(s,t_0)|$ indicates the
presence and intensity of the time scale $s$ mode in the time
series $x(t)$ at the moment of time $t_0$.

Fig.~\ref{fgr:WVTQuanatities},~{\it a} show the fragment of
electroencephalogram consisting of the laminar phases alternated
with turbulent ones (marked by gray rectangles) corresponding to
the epilepsy paroxysms, allocated by a method of expert
estimations. Fig.~\ref{fgr:WVTQuanatities},~{\it b} demonstrate
the result of the calculation wavelet surface $|W(s,t)|$
(\ref{eq:WVT_Phase}) for the fragment of electroencephalogram which
shown in Fig.~\ref{fgr:WVTQuanatities},~{\it a}. As one can see in
Fig.~\ref{fgr:WVTQuanatities},~{\it b}, wavelet surface structures
in the field of turbulent and laminar phases are essentially
different.
Since the laminar and turbulent phases are characterized by the
different distributions of the energy $E(s,t)=|W(s,t)|^2$ over
time scales $s$ (Fig.~\ref{fgr:WVTPowerDistrib}) the instantaneous
distribution of the energy of the wavelet spectrum averaged over
the characteristic interval ${s\in(s_1,s_2)}$ (${s_1=70}$~ms,
${s_2=110}$~ms) of the time scales
\begin{equation}
\langle E(t)\rangle=\int\limits_{s_1}^{s_2}E(s,t)\,ds
\end{equation}
may be used as a criterion which allows distinction between
different types of brain activity. Indeed, averaged energy of
turbulent phases $\langle E(t)\rangle$ exceeded a certain
threshold $\Delta E$, while laminar phases did not reach this
threshold level and $\langle E(t)\rangle$ is below $\Delta E$.
Therefore, we had a reliable threshold to extract laminar and
turbulent phases (see Fig.~\ref{fgr:WVTQuanatities},\,\textit{c}).
Duration of each phase was measured, and analysis was performed at the
presence of the ordered correspondence between phases. More
specifically, we studied distribution of laminar phases. Presence
of intermittent behavior was detected with analysis of EEG time
series.

\section{Results}
\label{sct:Results}

Continuous wavelet transform of EEG was used to establish
frontiers between normal and epileptic EEG. Hundredfold difference
in wavelet spectral energy was sufficient for reliable
identification periods of non-epileptic and epileptic EEG (see
Section~\ref{sct:Methods} and Fig.~\ref{fgr:WVTQuanatities},~{\it
a--c}). In general, automation of EEG analysis allowed
retrospective evaluation of expert assessments. It appeared that
expert estimates were completely congruent to the outcomes of
wavelet transform. Therefore, results of continuous wavelet
transform analysis were convenient and we may recommend using this
mathematical tool for others who needs to make a selection between
such kinds of the EEG activity.

Transition between normal and paroxysmal brain activity
(laminar and turbulent phases correspondingly) was not random, but
it revealed a clear intermittent behavior. More details on
selection of EEG representing laminar and turbulent phases could
be found in section~\ref{sct:Methods}, one essential note is that
normal EEG was best extrapolated to laminar phase and epileptic
activity (SWDs) --- to turbulent phase.

We found that the alternation of regimes of normal EEG and
paroxysmal activity (SWDs) fits well to on-off
intermittency~\cite{Platt:1993_intermittency,
Heagy:1994_intermittency} being observed in many
systems~\cite{Hramov:2005_IGS_EuroPhysicsLetters,%
Boccaletti:2000_IntermitLagSynchro}. This conclusion is based
mainly on the analysis of the distribution of lengths of
experimental laminar phases. Since in absence epilepsy
paroxysmal activity clearly depends on the circadian
periodicity~\cite{Luijtelaar:1988} (i.e. it follows the light-dark
cycle with night maximums), we analyzed datasets for light and
dark time periods separately.

\begin{figure}[b]
\centerline{\scalebox{0.6}{\includegraphics{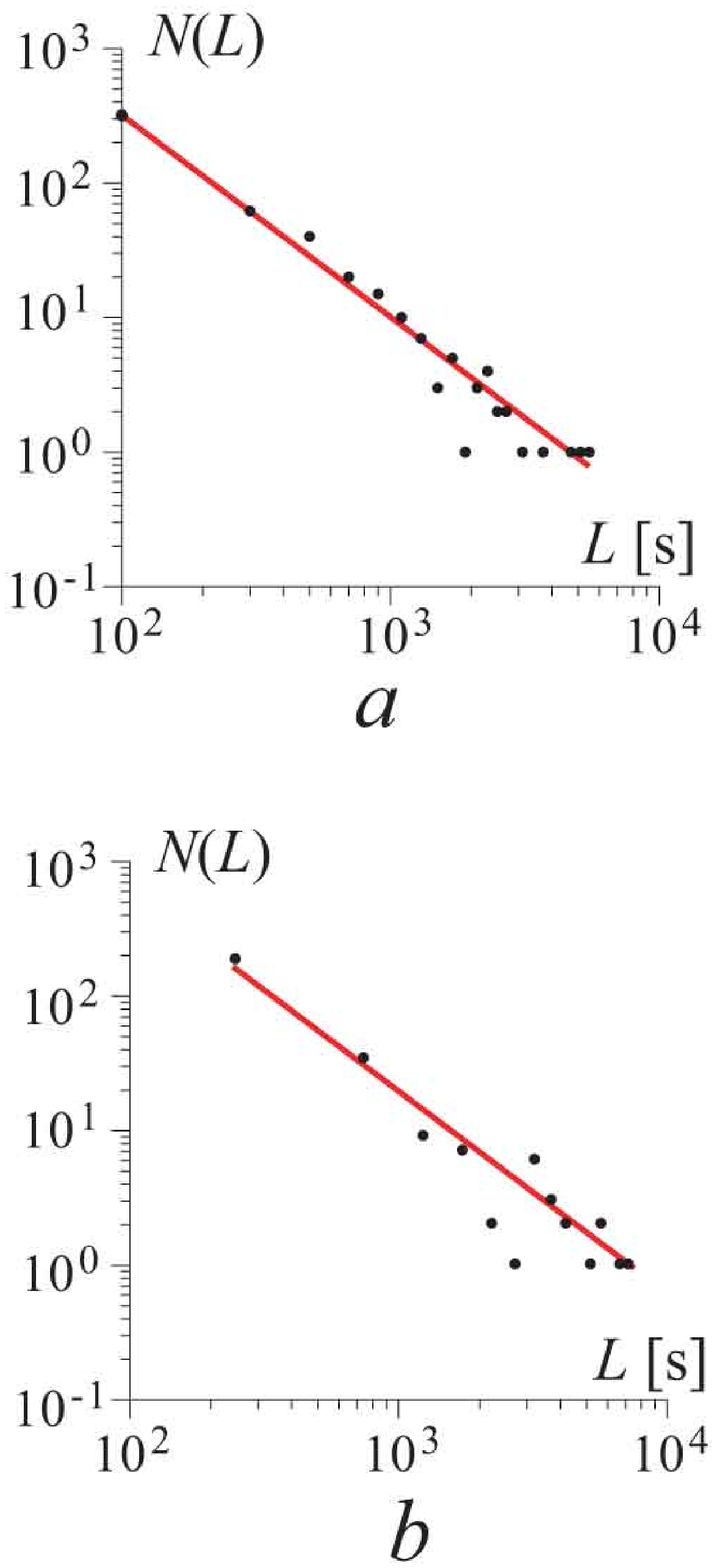}}}
\caption{The distribution of the length of the laminar phases of
electroencephalograms registered for (\textit{a}) dark and
(\textit{b}) light time periods. These distributions have been
plotted in the log-log scale. Black circles ($\bullet$) present
the experimental dots, the straight lines correspond to the power
law $N(L)\sim L^{-3/2}$} \label{fgr:N(L)Distributions}
\end{figure}

Fig.~\ref{fgr:N(L)Distributions},\,\textit{a,b} shows the
distributions $N(L)$ of laminar phase lengthes $L$ corresponding
to the dark and light periods, respectively. These distributions
$N(L)$ have been obtained from cumulative data registered during 4
days recording made in the same rat.

The distribution $N(L)$ of the laminar phases lengths $L$ for the
case of on-off intermittency is known to follow the power law
\begin{equation}
N(L)=\beta L^{\alpha},
\end{equation}
where $\alpha$ and $\beta$ are the parameters of power
distribution (exponent of power and normalization factors,
respectively), with ${\alpha=-3/2}$ (see, for
details~\cite{Platt:1993_intermittency,
Heagy:1994_intermittency}).

As one can see in Fig.~\ref{fgr:N(L)Distributions}, the
experimental distribution (shown by black circles) in the log-log
scale is close to the straight line corresponding to the power law
with the exponent $\alpha=-3/2$. The approximation of the
experimental data has been found by the least square method.
It is important, that on-off intermittent behavior has been diagnosed both
in dark and light periods, while the intensity of paroxysmal activity
was significantly different. As it is known, absence epilepsy paroxysms
in WAG/Rij rats are more abundant during the night time
compared to the day time~\cite{Luijtelaar:1986}. This explains why
the length of laminar phases showed lower average $\langle
L_{dark}\rangle$ compared to $\langle L_{light}\rangle$.
However, this decrease of $\langle L_{dark}\rangle$ do not breach
the power law with $\alpha=-3/2$. The same results were observed
for all analyzed data sets obtained in freely moving drug-naive
WAG/Rij rats of both sexes (6 females and 5 males). Based on that,
we conclude that the intermittent behavior found in time series of
paroxysmal activity in WAG/Rij rats is indeed in concordance with
regularities known for on-off intermittency.

\begin{figure}[tb]
\centerline{\scalebox{0.35}{\includegraphics{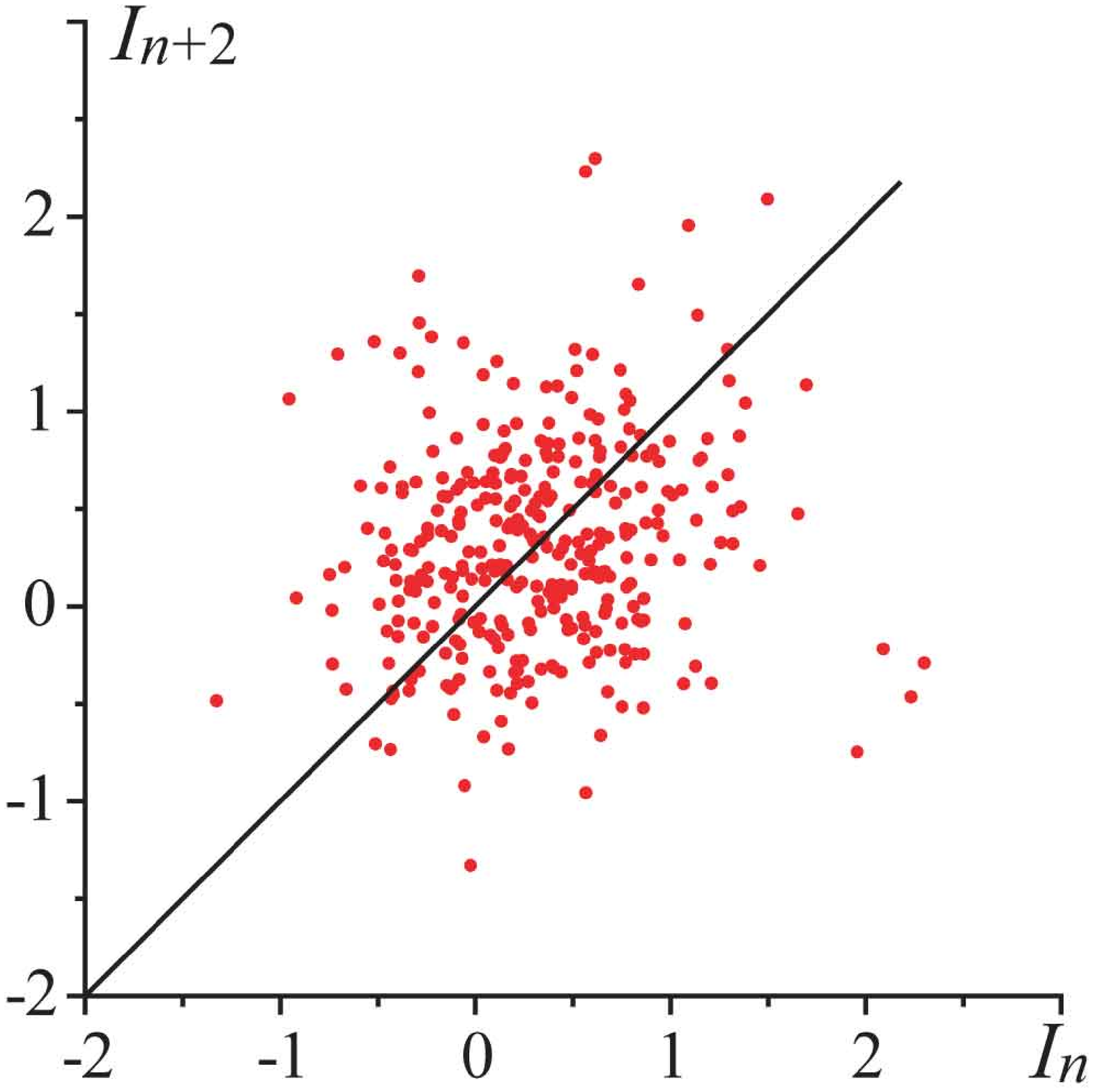}}}
\caption{The second return map obtained from the
electroencephalogram. Obviously, this map $I_{n+2}=F(I_n)$ does
not obey the function ${F(I)=(1+\varepsilon)I+bI^3}$ which should
be observed in the case of the intermittency--III}
\label{fgr:SecondReturnMap}
\end{figure}
The power law with the exponent $-3/2$ is known to be also
observed in the case of the intermittency--III for the
distribution of the short laminar phases (see,
e.g.,~\cite{Berge:1988_OrderInChaos}). Since the presence of the
intermittency--III may be revealed easily by the analysis of the
second return map in the same way as it was done
in~\cite{Dubois:1983_IntermittencyIII} we have checked whether
the experimental distribution $N(L)$ of laminar phases in the
electroencephalogram being under consideration is the
intermittency--III. In Fig.~\ref{fgr:SecondReturnMap} the second
return map obtained from the electroencephalogram is shown. One
can easily see, that this map can not be approximated by the cubic
polynomial ${f(I)=(1+\varepsilon)I+bI^3}$ taking place in the
case of the intermittency--III. Therefore, we make a decision that
the brain activity alternated between normal states and epilepsy
paroxysms is the on--off intermittency.

\begin{figure}[tb]
\centerline{\scalebox{0.3}{\includegraphics{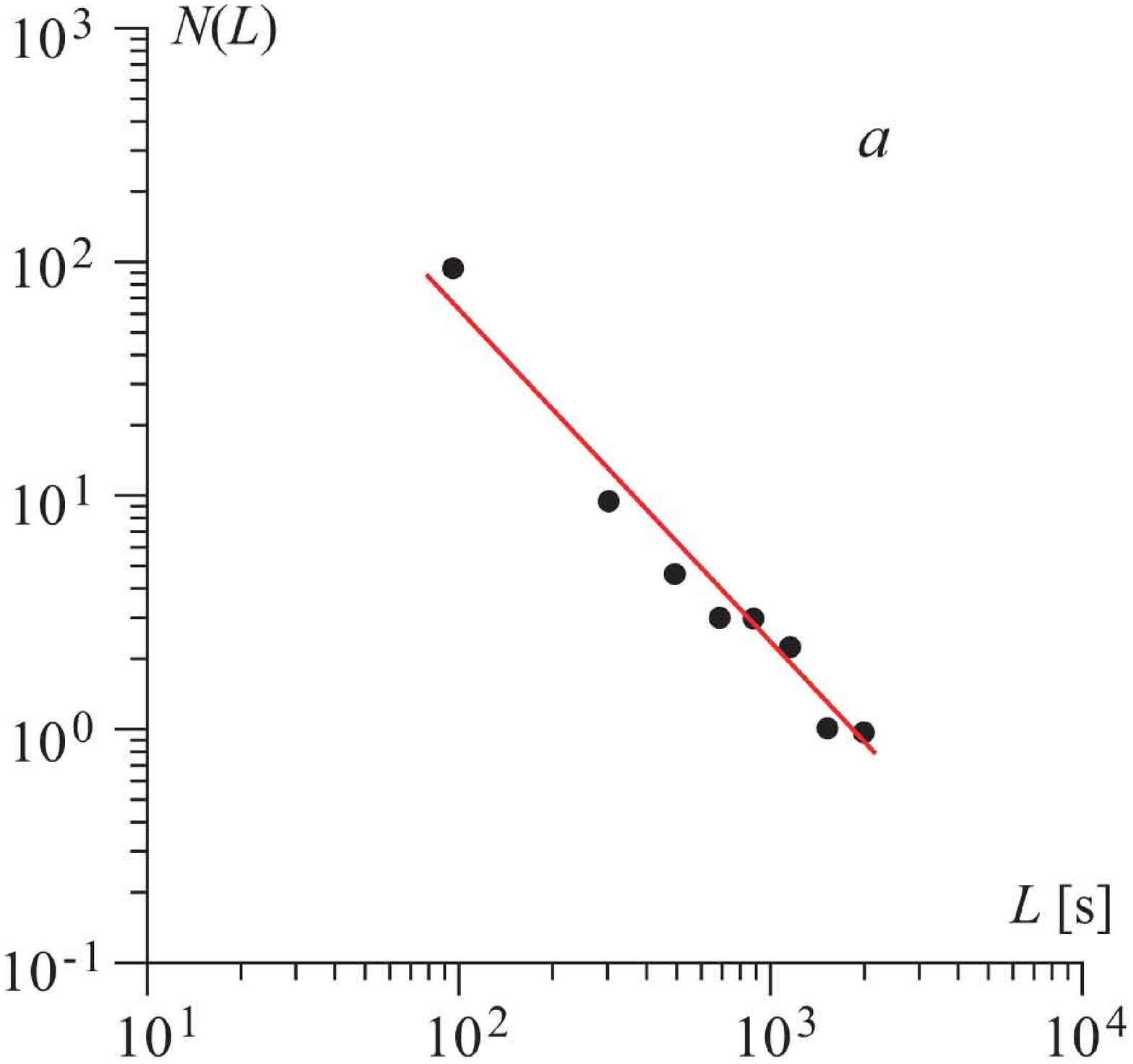}}}
\centerline{\scalebox{0.3}{\includegraphics{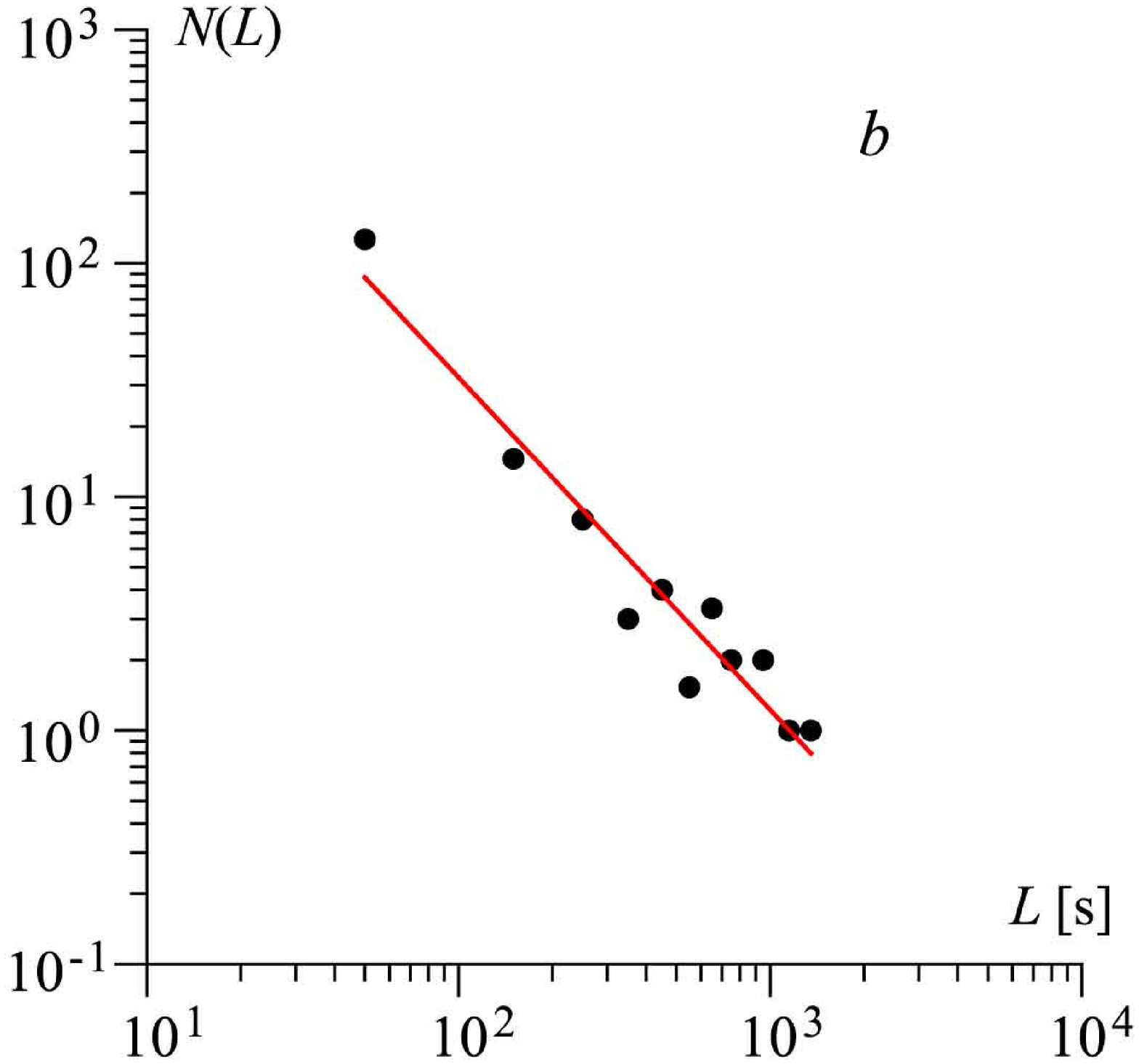}}}
\centerline{\scalebox{0.3}{\includegraphics{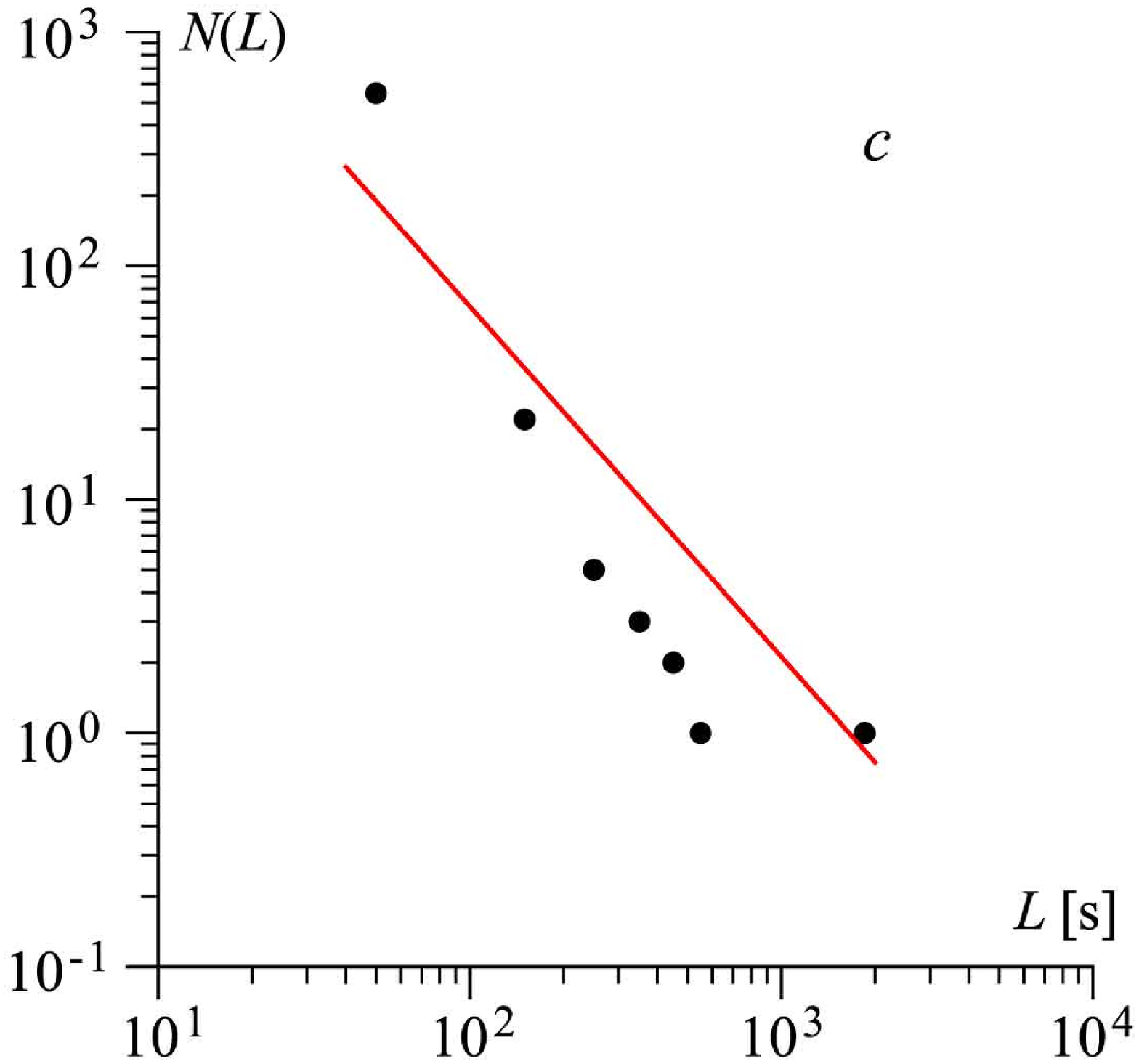}}}
 \caption{Distributions $N(L)$ of laminar phases lengths for EEG after i.p.
injections of vigabatrin in doses of (\textit{a}) $125$ mg,
(\textit{b}) $250$ mg and (\textit{c}) $500$ mg. These
distributions are based on shorter time series compared to
Fig.~\ref{fgr:N(L)Distributions}. Black circles ($\bullet$)
present the experimental dots, the straight lines correspond to
the power law $N(L)\sim L^{-3/2}$. Noteworthy that under
vigabatrin condition (with doses more than 500 mg), duration of
laminar phases did not follow the behavior of on-off intermittency
that was characteristic for drug-free state
Fig.~\ref{fgr:N(L)Distributions}} \label{fgr:Vigibatrin}
\end{figure}

We have found that in absence epilepsy, occurrence of EEG
paroxysms was not random, and spontaneous absence epilepsy could
be considered as an intermittent brain disorder. Mechanisms which
trigger epileptic seizures have still been poorly understood, yet
some external influences have been known to enhance or diminish
paroxysmal activity. In order to characterize how pharmacological
treatment influences dynamic properties of epileptic activity, we
injected a pro-absence drug, vigabatrin, in different doses. As it
is known, vigabatrin is good especially in the treatment of
generalized epilepsies~\cite{Appleton:1995_Vigibatrin}, but it has
an opposite effect in absence
epilepsy~\cite{Manning:2003_Pharmacology}. Vigabatrin causes
dose-dependent increase in number and duration of spike-wave
epileptic discharges~\cite{Vergnes:1984_Epilepsy}, which are
characteristic for absence epilepsy.
After injection of vigabatrin we tested intermittency of laminar
and turbulent phases for the presence of on-off pattern. It was
found that in lower doses ($125$, $250$ mg) vigabatrin changed
mean length of laminar phases, so that intervals with normal EEG
activity decreased. However, distortion of laminar phases after
injection of low-doses of vigabatrin did not disrupt to the power
law with the exponent $\alpha=-3/2$ (see
Fig.~\ref{fgr:Vigibatrin},\,\textit{a,b}). Pattern of on-off
intermittency in drug-injected animals was similar to that in
drug-free animals (Fig. 4). In general, low-doses of vibabatrin did
not alter global dynamic of epileptic activity. Injection of
vigabatrin in higher dose ($500$ mg) disrupted on-off
intermittency pattern of EEG activity (see
Fig.~\ref{fgr:Vigibatrin},\,\textit{c}), also lengths of laminar
phases were more randomly and did not correlate with $\alpha=-3/2$
exponent. This may imply that high-doses of vibabatrin influenced
transition from laminar to turbulent states so that intermittency
got another (probably, more complex) pattern compared to that in
drug-free animals and the animals injected with vigabatrin in lower
doses.

\section{Conclusion}
\label{sct:Conclusion}

In conclusion, we report that the behavior of the epileptic brain
in the case of spontaneous non-convulsive paroxysmal activity is
the on--off intermittency. We evaluate temporal aspects of brain
activity, in particular, relationship between two vitally
important states: spontaneous paroxysmal (epileptiform) activity
and non-paroxysmal (normal) activity. We state that paroxysmal
(turbulent) and non-paroxysmal (laminar) states exhibited on-off
intermittency with exponential rate $\alpha=-3/2$. Such a
regularity was typical for all experimental animals in the both light
and dark time periods. This clearly non-random behavior of EEG
dynamic in macro-scale may imply a powerful intrinsic mechanism
controlling occurrence of epileptic activity.

The obtained results may contribute to development of a more
general theory of epileptogenesis. It is known that one of the
mechanisms underlying on-off intermittency is a co-existence of
two different processes with one serving as a bifurcation
parameter for another (for details
see~\cite{Platt:1993_intermittency, Heagy:1994_intermittency}). To
this end, a feasibility of a paroxysm onset in the rats with
genetic an absence epilepsy can be modulated by some (yet unknown)
slow process, which in this way ``governs'' the occurrence and
elimination of epileptiform events. Taking into consideration that
in WAG/Rij rats the paroxysmal activity is aggravating with
age~\cite{Coenen:1987}, we can hypothesize some acceleration of
the second process with corresponded decrease of its time scope.
Hopefully, we will be able to deduce this putative governing
process from a special seria of experiments. Further
investigations can be of clinical interest because this way may
lead us to effective tools by which one can suppress
epileptogenesis.

\section*{Acknowledgements}

We thank Professors G.D.~Kuznetsova, Dmitry I. Trubetskov and Dr.
Svetlana V. Eremina for the support, Alexander A. Tyshchenko and
Olga I. Moskalenko for the help in the numerical calculations.
This work has been supported by the Russian Foundation for Basic
Research (grant 05--02--16286), the program of support of the
leading scientific schools (NSh--4167.2006.2) and by a special
grant ``Fundamental sciences for medicine'' from the Presidium of
Russian Academy of Sciences. A.E.H. and A.A.K. also thank
``Dynasty'' Foundation and ICFPM for financial support. A.E.H.
acknowledges support from CRDF, Grant No.~Y2--P--06--06.


\end{document}